\documentclass[aps,prev,twocolumn,preprintnumbers,floatfix,nofootinbib]{revtex4-1}
\pdfoutput=1
\usepackage[english]{babel}
\usepackage{graphicx,color}
\usepackage{bm}
\usepackage{times}
\usepackage{slashed}
\usepackage{multirow}
\usepackage[usenames,dvipsnames,svgnames,table]{xcolor}
\usepackage{adjustbox}
\usepackage{slashed}

\usepackage{booktabs}

\usepackage{amsthm}
\usepackage{amssymb,amsmath}
\usepackage{subfigure}
\usepackage{hyperref}
\usepackage[dvipsnames]{xcolor}
\definecolor{bluencs}{rgb}{0.0, 0.53, 0.74}
\definecolor{darkcyan}{rgb}{0.0, 0.55, 0.55}
\definecolor{hanblue}{rgb}{0.27, 0.42, 0.81}
\hypersetup{
     linktocpage=true,
     setpagesize=true,
     urlcolor=blue,
     citecolor=blue,
     linkcolor=blue, 
     menucolor=cyan,
     colorlinks=true, 
     filecolor=magenta,
     citebordercolor=blue,
     pdftitle={Investigation of Charged Higgs Bosons Production from Vector-Like $T$ Quark\\\vspace{0.1cm} Decays at  $e\gamma$ Collider},
     pdfsubject={Latex},
     pdfauthor={Bk},
    pdfkeywords={BSM},
    unicode=true
}
\usepackage[capitalize]{cleveref}

\newcommand{\be}{\begin{equation}}
\newcommand{\ee}{\end{equation}}
\newcommand{\bea}{\begin{eqnarray}}
\newcommand{\eea}{\end{eqnarray}}

\usepackage{xfrac}
 
\usepackage{float}
\usepackage{bbding,pifont}
\definecolor{brilliantrose}{rgb}{1.0, 0.33, 0.64}
\definecolor{lawngreen}{rgb}{0.49, 0.99, 0.0}
\definecolor{magenta}{rgb}{1.0, 0.0, 1.0}
\makeatletter
\let\old@float\@float
\def\@float{\let\centering\relax\old@float}
\makeatother
\begin{document}

\title{Investigation of Charged Higgs Bosons Production from Vector-Like $T$ Quark\\\vspace{0.1cm} Decays at  $e\gamma$ Collider}

\author{Rachid Benbrik}
\email{r.benbrik@uca.ac.ma}
\author{Mbark Berrouj}
\email{mbark.berrouj@ced.uca.ma}
\author{Mohammed Boukidi}
\email{mohammed.boukidi@ced.uca.ma}

\affiliation{Polydisciplinary Faculty, Laboratory of Fundamental and Applied Physics, Cadi Ayyad University, Sidi Bouzid, B.P. 4162, Safi, Morocco.}

\begin{abstract}

Within the extended framework of the Two-Higgs-Doublet Model Type II (2HDM-II), enhanced by a vector-like quark (VLQ) doublet $TB$, we present a comprehensive analysis of the process $e^{-}\gamma \rightarrow b\nu_{e}\bar{T}$ at future high-energy $e\gamma$ colliders, focusing on the decays $\bar{T} \rightarrow H^{-} \bar{b}$ and $H^{-} \rightarrow \bar{t}b$. Using current theoretical and experimental constraints, we calculate production cross sections for both unpolarized and polarized beams at center-of-mass energies of $\sqrt{s} = 2$ and 3 TeV, demonstrating that polarized beams significantly enhance detection prospects by increasing production rates. By analyzing kinematic distributions, we establish optimized selection criteria to effectively separate signal events from background. At $\sqrt{s} = 2$ TeV with an integrated luminosity of 1500 fb$^{-1}$, we find exclusion regions within $s_R^d \in [0.085, 0.16]$ for $m_T \in [1000, 1260]$ GeV and a discovery potential within $s_R^d \in [0.14, 0.17]$ for $m_T \in [1000, 1100]$ GeV, with these regions expanding to $s_R^d \in [0.05, 0.15]$ for $m_T \in [1000, 1340]$ GeV and $s_R^d \in [0.11, 0.17]$ for $m_T \in [1000, 1160]$ GeV at 3000 fb$^{-1}$. At $\sqrt{s} = 3$ TeV and 1500 fb$^{-1}$, we identify exclusion regions of $s_R^d \in [0.055, 0.135]$ for $m_T \in [1000, 1640]$ GeV and discovery regions of $s_R^d \in [0.09, 0.15]$ for $m_T \in [1000, 1400]$ GeV, which further expand to $s_R^d \in [0.028, 0.12]$ for $m_T \in [1000, 1970]$ GeV and $s_R^d \in [0.04, 0.122]$ for $m_T \in [1000, 1760]$ GeV at 3000 fb$^{-1}$. Our findings emphasize the increased detection potential at higher center-of-mass energies, particularly at 3 TeV compared to 2 TeV, with notable improvements when polarized beams are utilized. We also account for the effects of initial state radiation, beamstrahlung, and systematic uncertainties, which influence both exclusion and discovery prospects.
\end{abstract}

\maketitle

\section{Introduction}
\noindent 

The 2012 discovery of the Higgs boson by the ATLAS and CMS collaborations at the Large Hadron Collider (LHC) was a landmark achievement in particle physics, confirming the Higgs mechanism as the foundation for mass generation in the Standard Model (SM) \cite{ATLAS:2012yve, CMS:2012qbp}. Despite this success, unresolved theoretical issues in the SM’s scalar sector have driven the exploration of extended models. Among these, the Two-Higgs-Doublet Model (2HDM) stands out by introducing an additional Higgs doublet, thereby enriching the scalar sector with additional neutral ($H$, $A$) and charged ($H^\pm$) Higgs bosons \cite{Branco:2011iw, Draper:2020tyq}. These new scalar particles offer promising avenues for experimental investigation, including direct production and loop-induced processes at the LHC.

\noindent One compelling extension within the 2HDM framework is the incorporation of vector-like quarks (VLQs) \cite{Aguilar-Saavedra:2013qpa, Buchkremer:2013bha}. VLQs are heavy spin-1/2 fermions whose left- and right-handed components share identical electroweak quantum numbers, unlike their SM counterparts. This unique feature allows VLQs to interact with the Higgs sector independently of the standard electroweak symmetry-breaking dynamics. They can appear as singlets, doublets, or triplets under $SU(2)_L$, interacting with SM quarks through Yukawa couplings to the Higgs field \cite{delAguila:2000aa, Aguilar-Saavedra:2009xmz, DeSimone:2012fs, Lavoura:1992np, Buchkremer:2013bha, Alves:2023ufm}. VLQs are also predicted in several beyond-the-Standard-Model (BSM) frameworks, such as extra-dimensional models like Randall-Sundrum warped geometry \cite{Randall:1999ee, Carena:2007tn, Gopalakrishna:2011ef}, Grand Unified Theories (GUTs) based on $E_6$ symmetry \cite{Hewett:1988xc}, and models like Little Higgs \cite{Arkani-Hamed:2002ikv, Schmaltz:2005ky} and composite Higgs frameworks \cite{Dobrescu:1997nm, Chivukula:1998wd, He:2001fz, Hill:2002ap, Agashe:2004rs, Contino:2006qr, Barbieri:2007bh, Anastasiou:2009rv}. Additionally, VLQs can also appear in other theories, e.g., two-Higgs doublet models (2HDMs)~\cite{Arhrib:2024dou, Arhrib:2024tzm, Benbrik:2022kpo, Aguilar-Saavedra:2017giu, Dermisek:2019vkc, Badziak:2015zez, Angelescu:2015uiz, Arhrib:2016rlj}.

\noindent The inclusion of VLQs in the 2HDM framework opens new decay channels, such as $T/B \to H^\pm b/t$, $T/B \to H t/b$, and $T/B \to A t/b$, as well as $X/Y \to H^\pm t/b$. These decays significantly affect the branching ratios of VLQs, thereby modifying experimental constraints that have traditionally focused on SM-like decay channels, such as $Wq$, $Zq$, and $hq$. The presence of these new decay modes makes the 2HDM + VLQ scenario an attractive target for experimental searches, as it could lead to distinct signatures that have not been fully investigated at the LHC.

In terms of experimental searches, the ATLAS and CMS collaborations have conducted extensive investigations into VLQ production mechanisms, exploring both single and pair production channels \cite{ATLAS:2021ddx, ATLAS:2021gfv, ATLAS:2021ibc, CMS:2020ttz, CMS:2022yxp, CMS:2022tdo, CMS:2022fck}. However, these searches have yet to yield positive results, possibly because the VLQs predominantly decay into non-SM particles, which are not the primary focus of current searches. This observation underscores the necessity of exploring models like the 2HDM + VLQ, where VLQs are more likely to decay into additional Higgs states, potentially leading to distinctive signatures that have not been fully investigated at the LHC. Several studies have further investigated VLQs in this context \cite{Arhrib:2024nbj,Benbrik:2023xlo,  Kanemura:2015mxa, Chen:2017hak, Carvalho:2018jkq, Moretti:2016gkr, Prager:2017hnt, Prager:2017owg, Moretti:2017qby, Deandrea:2017rqp, Dermisek:2020gbr, Dermisek:2021zjd, Vignaroli:2015ama, Liu:2019jgp,Ghosh:2024boo,Ghosh:2023xhs}.

While hadron colliders like the LHC are excellent for studying a wide range of BSM physics, lepton colliders, such as the International Linear Collider (ILC) \cite{ILC:2007bjz, ILC:2013jhg} and the Compact Linear Collider (CLIC) \cite{Dannheim:2012rn, CLICDetector:2013tfe}, provide a cleaner experimental environment for precision measurements. These machines, designed to operate at center-of-mass energies ranging from 500 GeV to 3 TeV, enable $e^+e^-$, $\gamma\gamma$, and $e\gamma$ collisions via Compton scattering. Among these, $e\gamma$ collisions are particularly promising for the study of single vector-like top quark $T$ (VLT) production \cite{Yang:2020zjd, Shang:2020clm, Yang:2019xxr, Yang:2018fcx, Shang:2019zhh}, as they achieve higher production rates with polarized beams and reduce background interference.

In this work, we explore the production and decay of the VLT and the charged Higgs boson $H^\pm$ within the 2HDM+$TB$ doublet scenario at $e\gamma$ colliders \cite{Das:2023tna}. Specifically, we focus on the process $e^{-}\gamma \rightarrow b\nu_{e}\bar{T}$, followed by the decays $T \rightarrow H^+ b$ and $H^+ \rightarrow t\bar{b}$. Our analysis is conducted at center-of-mass energies of $\sqrt{s} = 2$ and 3 TeV, leveraging the enhanced production rates provided by polarized beams to assess the detectability of these particles.

The paper is structured as follows: In Section \ref{sec-F}, we introduce the 2HDM-II+VLQ framework and outline the simulation setup for the production and decay processes under study. Section \ref{sec-A} presents the numerical results for $\sqrt{s} = 2$ and 3 TeV. Finally, in Section \ref{sec:conclusion}, we summarize our findings and discuss their implications for future collider experiments.

\section{Framework}
\label{sec-F}
This section provides a concise overview of the 2HDM-II + VLQ framework. We begin by revisiting the well-known $\mathcal{CP}$-conserving scalar potential of the 2HDM, involving two Higgs doublets $(\Phi_1, \Phi_2)$ subject to a softly broken discrete $\mathbb{Z}_2$ symmetry, $\Phi_1 \to -\Phi_1$, which is softly violated by dimension-2 terms \cite{Branco:2011iw, Gunion:1989we}:

\begin{eqnarray} \label{pot}
\mathcal{V} &=& m^2_{11}\Phi_1^\dagger\Phi_1+m^2_{22}\Phi_2^\dagger\Phi_2
-\left(m^2_{12}\Phi_1^\dagger\Phi_2+{\rm h.c.}\right)
\nonumber \\
&&+\frac{1}{2}\lambda_1\left(\Phi_1^\dagger\Phi_1\right)^2
++\frac{1}{2}\lambda_2\left(\Phi_2^\dagger\Phi_2\right)^2 \nonumber \\
&& \qquad +\lambda_3\Phi_1^\dagger\Phi_1\Phi_2^\dagger\Phi_2
+\lambda_4\Phi_1^\dagger\Phi_2\Phi_2^\dagger\Phi_1
\nonumber \\
&&+\left[+\frac{1}{2}\lambda_5\left(\Phi_1^\dagger\Phi_2\right)^2+{\rm h.c.}\right].
\end{eqnarray}
All parameters in this potential are real. The two complex scalar doublets $\Phi_{1,2}$ can be rotated into a basis, $H_{1,2}$, where only one of them acquires a Vacuum Expectation Value (VEV). By employing the minimization conditions of the potential for EWSB, the 2HDM can be parametrized by seven independent quantities: $m_h$, $m_H$, $m_A$, $m_{H^\pm}$, $\tan\beta = v_2/v_1$, $\sin(\beta - \alpha)$, and the soft-breaking parameter $m^2_{12}$

To suppress tree-level Flavor Changing Neutral Currents (FCNCs), the 2HDM admits four distinct Yukawa configurations, depending on how the $\mathbb{Z}_2$ symmetry is extended to the fermion sector\footnote{This work specifically focuses on the Type-II 2HDM within the alignment limit \cite{Draper:2020tyq}, where the lightest neutral Higgs boson is identified with the discovered 125 GeV state.}. These configurations are: Type-I, where $\Phi_2$ couples to all fermions; Type-II, where $\Phi_2$ couples to up-type quarks and $\Phi_1$ to down-type quarks and charged leptons; Type-Y (Flipped), where $\Phi_2$ couples to up-type quarks and charged leptons, and $\Phi_1$ to down-type quarks; and Type-X (Lepton Specific), where $\Phi_2$ couples to quarks, and $\Phi_1$ to charged leptons.

Next, we introduce the VLQ component of the model. The gauge-invariant interactions between the new VLQs and SM particles arise from renormalizable couplings, with the VLQ representations specified as follows:
\begin{align}
& T_{L,R}^0 \, && \text{(singlets)} \,, \notag \\
& (X\,T^0)_{L,R} \,, \quad (T^0\,B^0)_{L,R} \, && \text{(doublets)} \,, \notag \\
& (X\,T^0\,B^0)_{L,R} \,, \quad (T^0\,B^0\,Y)_{L,R}  && \text{(triplets)} \,.
\end{align}
In this context, the superscript zero distinguishes weak eigenstates from mass eigenstates. The electric charges of the VLQs are $ Q_T = 2/3 $, $ Q_B = -1/3 $, $ Q_X = 5/3 $, and $ Q_Y = -4/3 $, with $T$ and $B$ sharing the same electric charges as the SM top and bottom quarks, respectively.

The physical up-type quark mass eigenstates may, in general, contain non-zero $Q_{L,R}^0$ (with $Q$ being the VLQ field) components, when new fields $T_{L,R}^0$ of charge $2/3$ and non-standard isospin assignments are added to the SM. This situation leads to a deviation in their couplings to the $Z$ boson. Atomic parity violation experiments and the measurement of $R_c$ at LEP impose constraints on these deviations for the up and charm quarks, significantly stronger than those for the top quark.
In the Higgs basis, the Yukawa Lagrangian contains the following terms:
\begin{equation}
-\mathcal{L} \,\, \supset  \,\, y^u \bar{Q}^0_L \tilde{H}_2 u^0_R +  y^d \bar{Q}^0_L H_1 d^0_R + M^0_u \bar{u}^0_L u^0_R  + M^0_d \bar{d}^0_L d^0_R + \rm {h.c}.
\end{equation}
Here, $u_R$ actually runs over $(u_R, c_R, t_R, T_R)$ and $d_R$ actually runs over $(d_R, s_R, b_R, B_R)$. 

We now turn to the mixing of the new partners to the third generation, $y_u$ and $y_d$, which are $3\times 4$ Yukawa matrices. In fact, in the light of the above constraints, 
it is very reasonable to assume that only the top quark $t$  ``mixes'' with $T$.
In this case, the $2 \times 2$ unitary matrices $U_{L,R}^u$ define the relation between the charge $2/3$ weak and mass eigenstates:
\begin{eqnarray}
\left(\! \begin{array}{c} t_{L,R} \\ T_{L,R} \end{array} \!\right) &=&
U_{L,R}^u \left(\! \begin{array}{c} t^0_{L,R} \\ T^0_{L,R} \end{array} \!\right)
\nonumber \\
&=& \left(\! \begin{array}{cc} \cos \theta_{L,R}^u & -\sin \theta_{L,R}^u e^{i \phi_u} \\ \sin \theta_{L,R}^u e^{-i \phi_u} & \cos \theta_{L,R}^u \end{array}
\!\right)
\left(\! \begin{array}{c} t^0_{L,R} \\ T^0_{L,R} \end{array} \!\right) \,.
\label{ec:mixu}
\end{eqnarray}
In contrast to the up-type quark sector, the addition of new fields $B_{L,R}^0$ of charge $-1/3$ in the down-type quark sector results in four mass eigenstates $d,s,b,B$.
Measurements of $R_b$ at LEP set constraints on the $b$ mixing with the new fields that are stronger than for mixing with the lighter quarks $d,s$. 

In this case, then,  $2 \times 2$ unitary matrices $U_{L,R}^d$ define the dominant $b-B$ mixing as 
\begin{eqnarray}
\left(\! \begin{array}{c} b_{L,R} \\ B_{L,R} \end{array} \!\right)
&=& U_{L,R}^d \left(\! \begin{array}{c} b^0_{L,R} \\ B^0_{L,R} \end{array} \!\right)
\nonumber \\
&=& \left(\! \begin{array}{cc} \cos \theta_{L,R}^d & -\sin \theta_{L,R}^d e^{i \phi_d} \\ \sin \theta_{L,R}^d e^{-i \phi_d} & \cos \theta_{L,R}^d \end{array}
\!\right)
\left(\! \begin{array}{c} b^0_{L,R} \\ B^0_{L,R} \end{array} \!\right) \,.
\label{ec:mixd}
\end{eqnarray}
In the weak eigenstate basis, diagonalizing the mass matrices leads to the Lagrangian for the third generation and heavy quark mass terms:

\begin{eqnarray}
\mathcal{L}_\text{mass} & = & - \left(\! \begin{array}{cc} \bar t_L^0 & \bar T_L^0 \end{array} \!\right)
\left(\! \begin{array}{cc} y_{33}^u \frac{v}{\sqrt 2} & y_{34}^u \frac{v}{\sqrt 2} \\ y_{43}^u \frac{v}{\sqrt 2} & M^0 \end{array} \!\right)
\left(\! \begin{array}{c} t^0_R \\ T^0_R \end{array}
\!\right) \notag \\
& & - \left(\! \begin{array}{cc} \bar b_L^0 & \bar B_L^0 \end{array} \!\right)
\left(\! \begin{array}{cc} y_{33}^d \frac{v}{\sqrt 2} & y_{34}^d \frac{v}{\sqrt 2} \\ y_{43}^d \frac{v}{\sqrt 2} & M^0 \end{array} \!\right)
\left(\! \begin{array}{c} b^0_R \\ B^0_R \end{array}
\!\right) +\text{h.c.}
\label{ec:Lmass}
\end{eqnarray}
where $M^0$ represents a bare mass term, $y_{ij}^q$ are the Yukawa couplings for $q = u, d$, and $v = 246$ GeV is the Higgs VEV in the SM. By applying standard diagonalization techniques, the mixing matrices are determined by:
\begin{equation}
U_L^q \, \mathcal{M}^q \, (U_R^q)^\dagger = \mathcal{M}^q_\text{diag} \,,
\label{ec:diag}
\end{equation}
where $\mathcal{M}^q$ corresponds to the mass matrices in Eq.~(\ref{ec:Lmass}), and $\mathcal{M}^q_\text{diag}$ represents the diagonalized mass matrices. To ensure consistency, the $2 \times 2$ mass matrix reduces to the SM quark mass term if either the $T$ or $B$ quarks are absent.

It is also important to note that in representations containing both $T$ and $B$ quarks, the bare mass term remains identical for both up- and down-type quark sectors. For singlets and triplets, $y_{43}^q = 0$, while for doublets, $y_{34}^q = 0$. The mixing angles in the left- and right-handed sectors are not independent parameters. Through bi-unitary diagonalization of the mass matrix in Eq.~(\ref{ec:diag}), we obtain the following relations:
\begin{eqnarray}
\tan 2 \theta_L^q & = & \frac{\sqrt{2} |y_{34}^q| v M^0}{(M^0)^2-|y_{33}^q|^2 v^2/2 - |y_{34}^q|^2 v^2/2} \quad \text{(singlets, triplets)} \,, \notag \\
\tan 2 \theta_R^q & = & \ \frac{\sqrt{2}  |y_{43}^q| v M^0}{(M^0)^2-|y_{33}^q|^2 v^2/2 - |y_{43}^q|^2 v^2/2} \quad \text{(doublets)} \,,
\label{ec:angle1}
\end{eqnarray}

with the corresponding angle relations:
\begin{eqnarray}
\tan \theta_R^q & = & \frac{m_q}{m_Q} \tan \theta_L^q \quad \text{(singlets, triplets)} \,, \notag \\
\tan \theta_L^q & = & \frac{m_q}{m_Q} \tan \theta_R^q \quad \text{(doublets)} \,,
\label{ec:rel-angle1}
\end{eqnarray}
where $(q,m_q,m_Q) = (u,m_t,m_T)$ and $(d,m_b,m_B)$. Notably, one of the mixing angles always dominates, particularly in the down-type quark sector (for further details on this Lagrangian formalism, refer to Ref.~\cite{Arhrib:2024tzm}). For convenience, the superscripts $u$ and $d$ are omitted whenever the mixing occurs exclusively in the up- or down-type quark sector. Additionally, we use shorthand notations such as $s_{L,R}^{u,d} \equiv \sin \theta_{L,R}^{u,d}$ and $c_{L,R}^{u,d} \equiv \cos \theta_{L,R}^{u,d}$.

Building on the analysis in Refs~\cite{Arhrib:2024tzm,Benbrik:2022kpo}, our study is strategically focused on the 2HDM-II + $TB$ doublet scenario. This choice is motivated by our objective to simultaneously probe both the VLT quark and the charged Higgs boson. To achieve this, we sought a configuration where the production of the charged Higgs boson via VLT decays is maximized. Among all possible representations, the $TB$ doublet stands out as it offers the highest production rate, with the charged Higgs boson being produced in nearly 100\% of VLT decays. The expression for the decay width in the 2HDM-II + $TB$ scenario is as follows\footnote{Detailed analytical calculations of the couplings are provided in the Appendix of Ref.~\cite{Benbrik:2022kpo}.}:
\begin{align}
\Gamma(T \to H^+ b) & = \frac{g^2 }{64 \pi} 
\frac{m_T}{M_W^2} \lambda(m_T,m_b,M_{H^\pm})^{1/2}
\nonumber \\
&\times\left\{ (|Z_{Tb}^L|^2 \cot^2 \beta + |Z_{Tb}^R|^2 \tan^2 \beta )\right. \notag \\
& \left. \times  \left[1+r_b^2 - r_{H^\pm}^2 \right]  + 4 r_b \mathrm{Re}(Z_{Tb}^L) Z_{Tb}^{R*} \right\} \,.
\label{ec:GammaT}
\end{align}
Here, $r_x = m_x / m_T$, where $x$ refers to one of the decay products, and the function $\lambda(x,y,z)$ is defined as:
\begin{equation}
\lambda(x,y,z) \equiv (x^4 + y^4 + z^4 - 2 x^2 y^2 
- 2 x^2 z^2 - 2 y^2 z^2) \,,
\end{equation}%
and \begin{eqnarray}
\begin{array}{cc}
Z^L_{Tb}=c_L^d s_L^u e^{- i\phi_u}  +  ( s_L^u{}^2 -  s_R^u{}^2  )\frac{s_L^d}{c_L^u},\\Z^R_{Tb}= \frac{m_b}{ m_T} \left[    c_L^d s_L^u +  (s_R^d{}^2   -  s_L^d{}^2 ) \frac{c_L^u}{s_L^d}     \right]
\end{array}
\end{eqnarray}
\section{Theoretical and Experimental Bounds}
\label{sec-A}
In this section, we outline the constraints used to validate our results.
\begin{itemize}
	\item \textbf{Unitarity} constraints require the $S$-wave component of the various
	(pseudo)scalar-(pseudo)scalar, (pseudo)scalar-gauge boson, and gauge-gauge bosons scatterings to be unitary
	at high energy ~\cite{Kanemura:1993hm}.
	\item \textbf{Perturbativity} constraints impose the following condition on the quartic couplings of the scalar potential: $|\lambda_i|<8\pi$ ($i=1, ...5$)~\cite{Branco:2011iw}.    
	\item \textbf{Vacuum stability} constraints require the potential to be bounded from below and positive in any arbitrary
	direction in the field space, as a consequence, the $\lambda_i$ parameters should satisfy the conditions as~\cite{Barroso:2013awa,Deshpande:1977rw}:
	\begin{align}
	\lambda_1 > 0,\quad\lambda_2>0, \quad\lambda_3>-\sqrt{\lambda_1\lambda_2} ,\nonumber\\ \lambda_3+\lambda_4-|\lambda_5|>-\sqrt{\lambda_1\lambda_2}.\hspace{0.5cm}
	\end{align} 
	\item \textbf{Constraints from EWPOs}, implemented through the oblique parameters\footnote{A comprehensive discussion on EWPO contributions in VLQs can be found in~\cite{Benbrik:2022kpo, Abouabid:2023mbu}}, $S$ and $T$ ~\cite{Grimus:2007if},  require that, for a parameter point of our
	model to be allowed, the corresponding $\chi^2(S^{\mathrm{2HDM\text{-}II}}+S^{\mathrm{VLQ}},~T^{\mathrm{2HDM\text{-}II}}+T^{\mathrm{VLQ}})$ is within 95\% Confidence Level (CL) in matching the global fit results \cite{Molewski:2021ogs}:
	\begin{align}
	S= 0.05& \pm 0.08,\quad T = 0.09 \pm 0.07,\nonumber \\ &\rho_{S,T} = 0.92 \pm 0.11. 
	\end{align}
	Note that unitarity, perturbativity, vacuum stability, as well as $S$ and $T$ constraints, are enforced through the public code  \texttt{2HDMC-1.8.0}\footnote{The code has been adjusted to include new VLQ couplings, along with the integration of analytical expressions for $S_{VLQs}$ and $T_{VLQs}$ outlined in Ref.~\cite{Arhrib:2024tzm}.} \cite{Eriksson:2009ws}.
	\item \textbf{Constraints from the SM-like Higgs-boson properties}  are taken into account by using \texttt{HiggsSignal-3} \cite{Bechtle:2020pkv,Bechtle:2020uwn} via \texttt{HiggsTools} \cite{Bahl:2022igd}. We require that the relevant quantities (signal strengths, etc.) satisfy $\Delta\chi^2=\chi^2-\chi^2_{\mathrm{min}}$ for these measurements at 95\% CL ($\Delta\chi^2\le6.18$).
	\item\textbf{Constraints from direct searches at colliders}, i.e., LEP, Tevatron, and LHC, are taken at the 95\% CL and are tested using \texttt{HiggsBouns-6}\cite{Bechtle:2008jh,Bechtle:2011sb,Bechtle:2013wla,Bechtle:2015pma} via \texttt{HiggsTools}. Including the most recent searches for neutral and charged scalars.
	
	\item {\bf Constraints from $b\to s\gamma$}:  As established in \cite{Benbrik:2022kpo}, the Type-II 2HDM typically requires the charged Higgs boson mass to exceed 580 GeV to satisfy $b \to s\gamma$ constraint. The introduction of VLQs into the 2HDM can relax this stringent requirement, particularly when larger mixing angles are considered. However, restrictions from EWPOs limit the extent of these mixing angles, permitting charged Higgs masses below 580 GeV. In typical scenarios, the charged Higgs mass remains around 580 GeV for the 2HDM+$T$ singlet and approximately 360 GeV or higher for the 2HDM-II + $TB$ doublet, though the potential for lower masses exists under these relaxed conditions.

	\item \textbf{LHC direct search constraints for VLQs}: The current LHC limits primarily target the SM decay modes of VLT, such as $T \to Wb$, $ht$, and $Zt$. In our scenario, however, additional decay channels like $T \to H^\pm b$, $Ht$, and $At$ become relevant, which may influence these limits. To ensure accuracy, we incorporated the latest LHC constraints \cite{ATLAS:2022ozf,ATLAS:2023pja,ATLAS:2023bfh,ATLAS:2024xne,CMS:2022yxp,CMS:2023agg} into our analysis. We implemented a stringent condition where only parameter points meeting the criterion $r={\sigma_{\mathrm{theo}}}/{\sigma_{\mathrm{obs}}^{\mathrm{LHC}}} < 1$ were retained, signifying that points with $r > 1$ are excluded at the 95\% CL.	
\end{itemize}

\section{Analysis and Simulation}
\label{sec3}

Our analysis is focused on the 2HDM+$TB$ scenario, motivated by the dominance of the charged Higgs boson production as the primary decay channel for the VLT within this framework. Specifically, the branching ratio for the $T \rightarrow H^+b$ decay can approach nearly 100\% for large $m_T$ values, making it the predominant decay mode. This distinguishes the 2HDM+$TB$ model from other VLQ representations, where alternative decay channels may be more competitive. For further details on this behavior, we refer to Ref.\cite{Benbrik:2022kpo}. As illustrated in Fig.\ref{fig2}, the $T \rightarrow H^+b$ decay mode becomes dominant for $m_T$ values above 800 GeV.

\begin{table}[ht!]
	\centering
	{\renewcommand{\arraystretch}{1.} 
		{\setlength{\tabcolsep}{0.15\columnwidth} 
			\begin{tabular}{cc}\hline  
				\hline
				Parameter  & Range \\
				\hline			
				$m_h$   & $125.09$ GeV \\
				$m_A$  & [$400$, $800$] GeV \\
				$m_H$  & [$400$, $800$] GeV \\
				$m_{H^\pm}$  & [$400$, $800$] GeV \\
				$t_\beta$ & [$1$, $20$] \\
				$s_{\beta-\alpha}$&1\\
				$m^2_{12}$ &$m_A^2s_\beta c_\beta$\\
				$m_{T}$   & [$500$, $2000$] GeV \\	
				$s_L^{u}$  & [$-0.8$, $0.8$] \\
				$s_R^{d}$  & [$-0.8$, $0.8$] \\
				\hline\hline
			\end{tabular}	
			\caption{Parameter ranges for 2HDM+$TB$ explored in the analysis.}\label{tab:Tab1}}}	
\end{table}
\begin{figure}[ht]
	\centering
	\includegraphics[height=7cm,width=0.94\columnwidth]{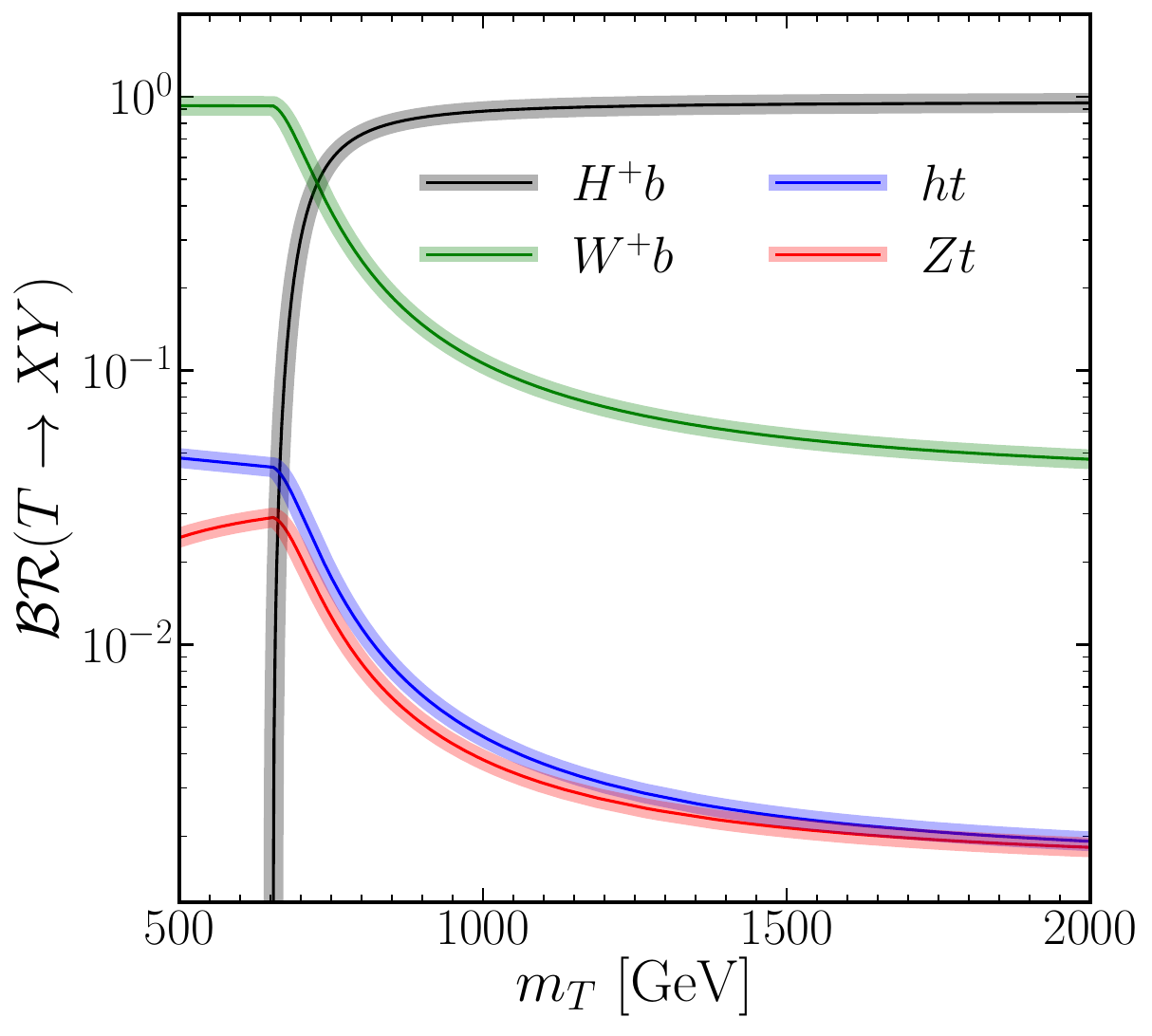}
	\caption{Branching ratios of different decay channels of the VLT as a function of $m_T$. The black line corresponds to $T \rightarrow H^+b$, green to $T \rightarrow W^+b$, blue to $T \rightarrow ht$, and orange to $T \rightarrow Zt$. Parameters used: $m_H=600.26$ GeV, $m_A=	595.24$ GeV, $m_{H^\pm}= 658.07$ GeV, $\tan\beta=6$, $s^u_R=0.05$ and $s^d_R=0.1$.}
	\label{fig:BR}
\end{figure}
To explore the parameter space systematically, we conducted a scan over the ranges specified in Table \ref{tab:Tab1}, leading to the identification of three benchmark points (BPs) that satisfy all relevant theoretical and experimental constraints. The selected benchmarks, corresponding to $m_T = 1010$, 1321, and 1605 GeV, are detailed in Table \ref{tab:Tab2}.

\begin{table}[H]
	\begin{center}
		\setlength{\tabcolsep}{12pt}
		\renewcommand{\arraystretch}{1.1}
		\begin{adjustbox}{max width=\textwidth}		
			\begin{tabular}{lccc}
				\hline\hline
				Parameters &       BP$_1$ &       BP$_2$ &       BP$_3$ \\
				\hline
				$m_h$   &   125.09 &   125.09&   125.09   \\
				$m_H$ &  594.32 &   616.79 &  627.84  \\
				$m_A$  &  586.81  & 617.02 &   622.33 \\
				$m_{H\pm}$ &  608.38  &   607.42 &  655.57  \\
				$t_\beta$ &    4.4 &  4.78 &     4.89  \\
				$m_T$   & 1010.02   &  1321.87 & 1605.04  \\
				$m_B$     & 1029.20 &  1340.28 & 1619.83  \\
				$s^u_L$    &    0.0038  &  0.0059 &   0.0041  \\
				$s^d_L$     &    0.0009  & 0.0006 &    0.0004 \\
				$s^u_R$     &    0.0223  &  0.0449&   0.0383  \\
				$s^d_R$   &    0.1934    &  0.1709 &    0.1400 \\ \hline\hline	
			\end{tabular}
		\end{adjustbox}
	\end{center}
	\caption{The description of our BPs. Masses are in GeV.}\label{tab:Tab2}
\end{table}
\begin{figure}[H]
	\centering
	\includegraphics[height=7cm,width=0.9\columnwidth]{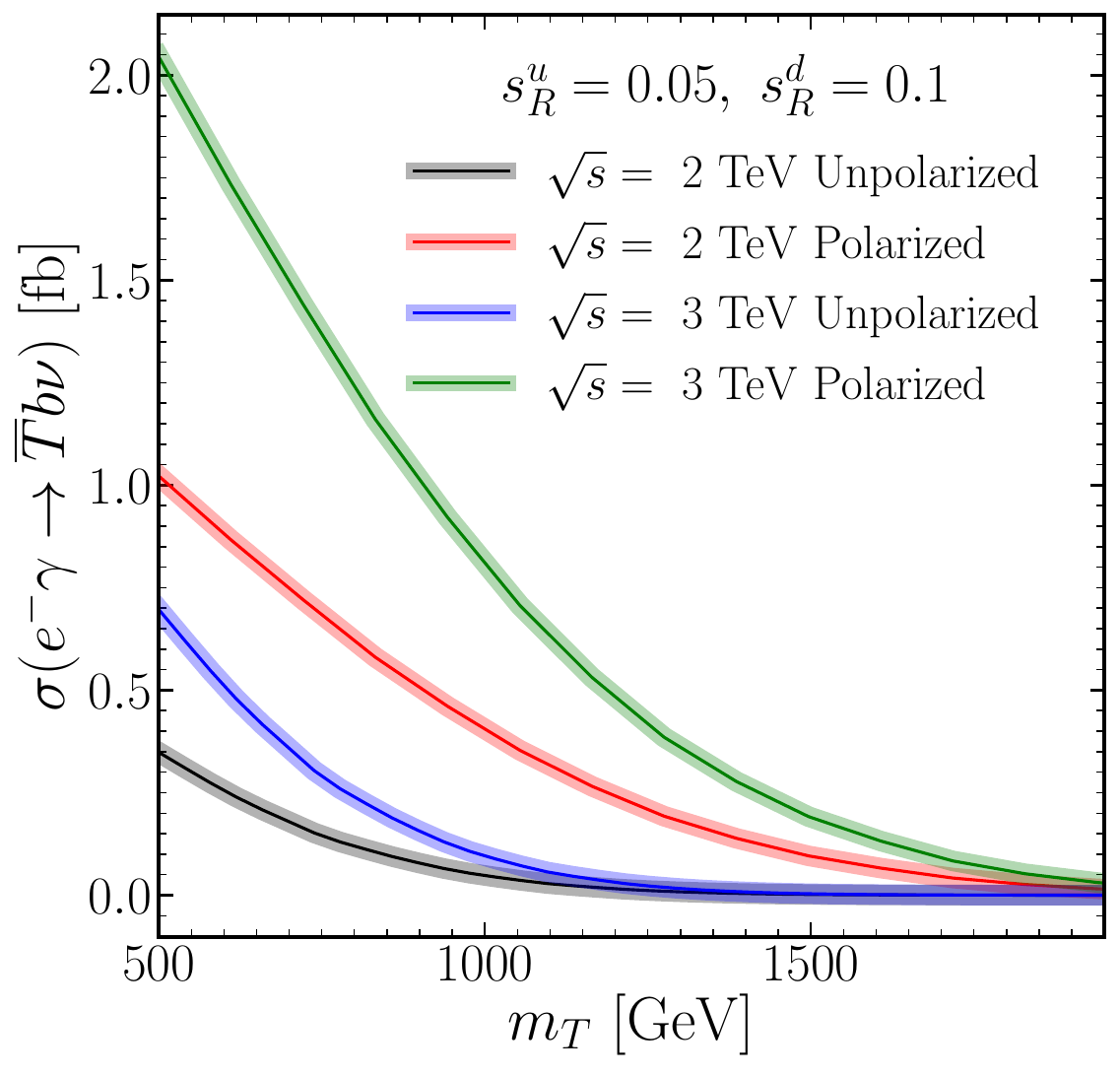}
	\caption{Cross section $\sigma(e^-\gamma\to \overline{T}b \nu)$ as a function of $m_T$) for different center-of-mass energies $\sqrt{s}$ with fixed parameters: $s^u_R=0.05$, and $s^d_R=0.1$.}
	\label{fig:XS}
\end{figure}

The leading-order Feynman diagrams for the single production of the VLT at an $e\gamma$ collider are depicted in Fig.~\ref{fig2}. The full production and decay chain considered in this study is as follows:

$\gamma e^- \rightarrow (\bar{T} \rightarrow H^{-} \bar{b} \rightarrow \bar{t} b \bar{b} \rightarrow W b b \bar{b} \rightarrow l^{+}\nu_{l} b b \bar{b}) b \nu_{e}.$

In this process, the charged Higgs boson $H^-$ decays into $\bar{t}b$, with the top quark subsequently decaying leptonically via $W^- \to l^- \nu_{l}$. As a result, the final state consists of one isolated charged lepton ($l = e,~\mu$), four b-jets, and missing transverse energy ($\slashed{E}_T$).

\begin{figure}[H]
	\begin{center}
		\begin{minipage}{\textwidth}
			\centering
			\includegraphics[height=3.cm,width=4cm]{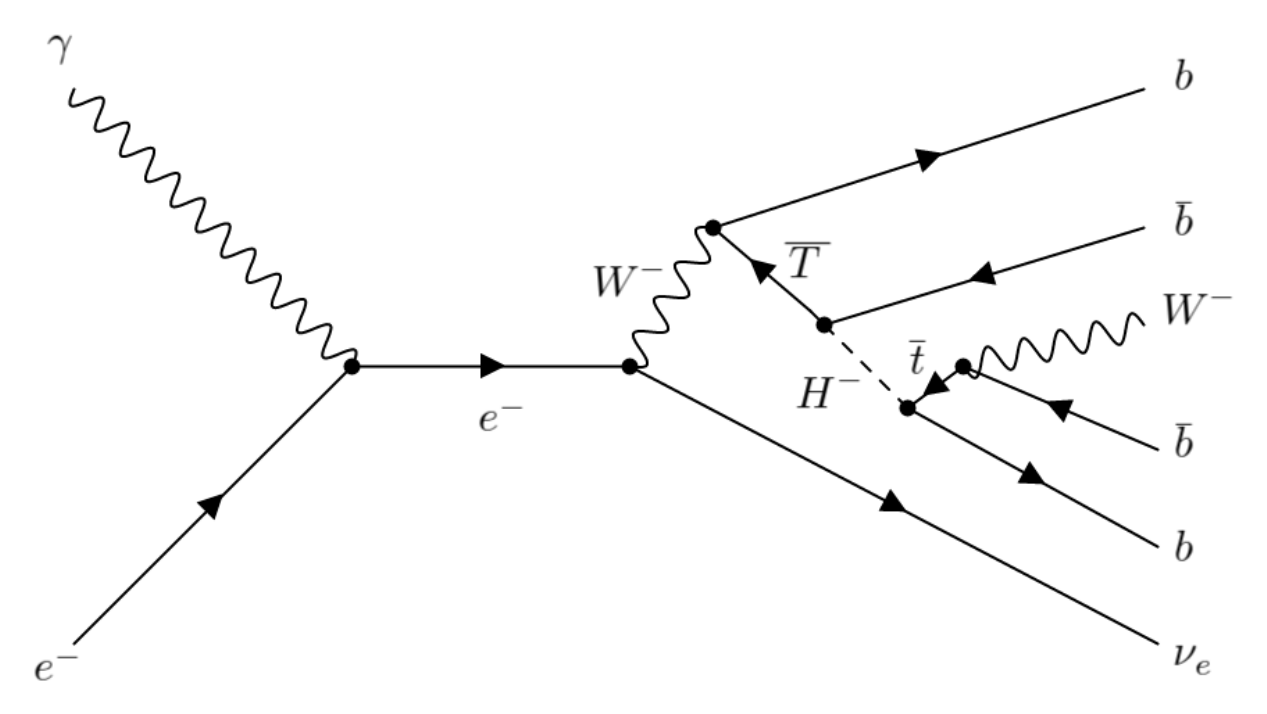}
			\includegraphics[height=3.cm,width=4cm]{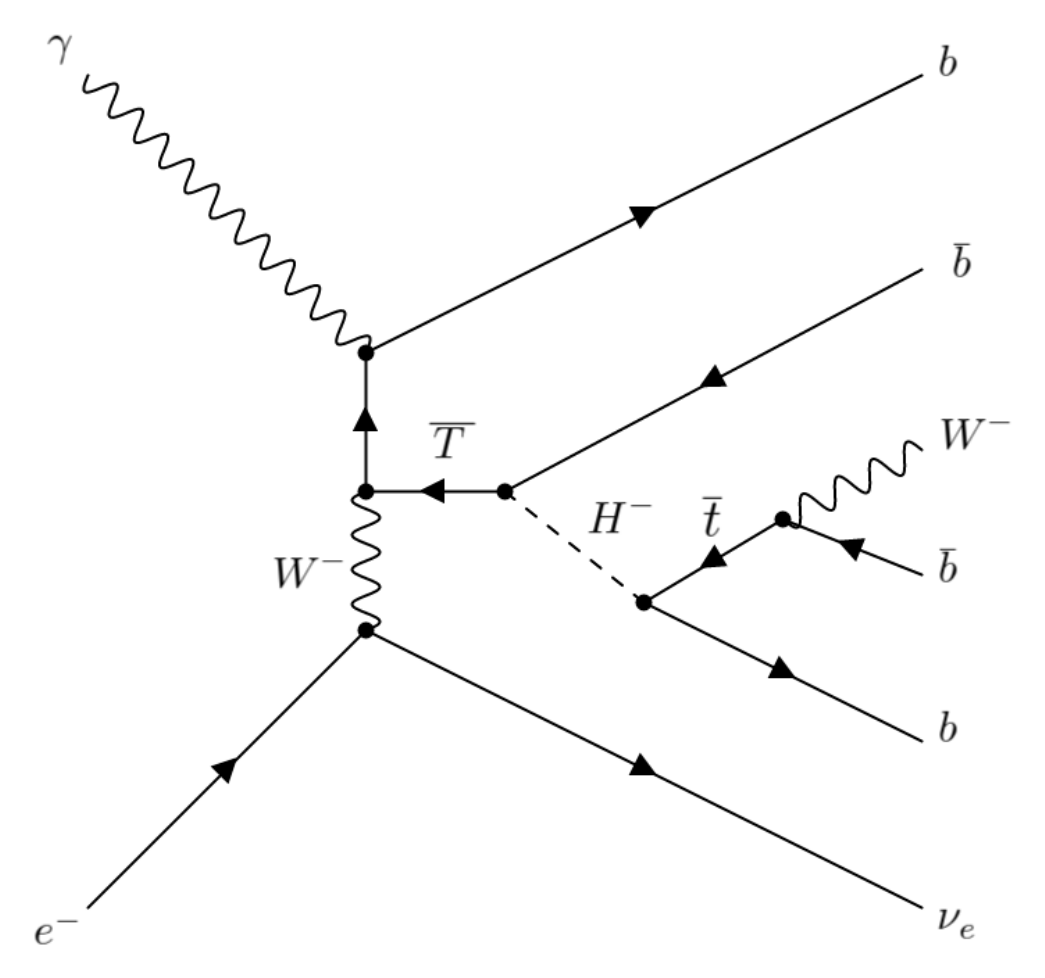}
		\end{minipage}
		\begin{minipage}{\textwidth}
			\centering
			\includegraphics[height=3.8cm,width=4cm]{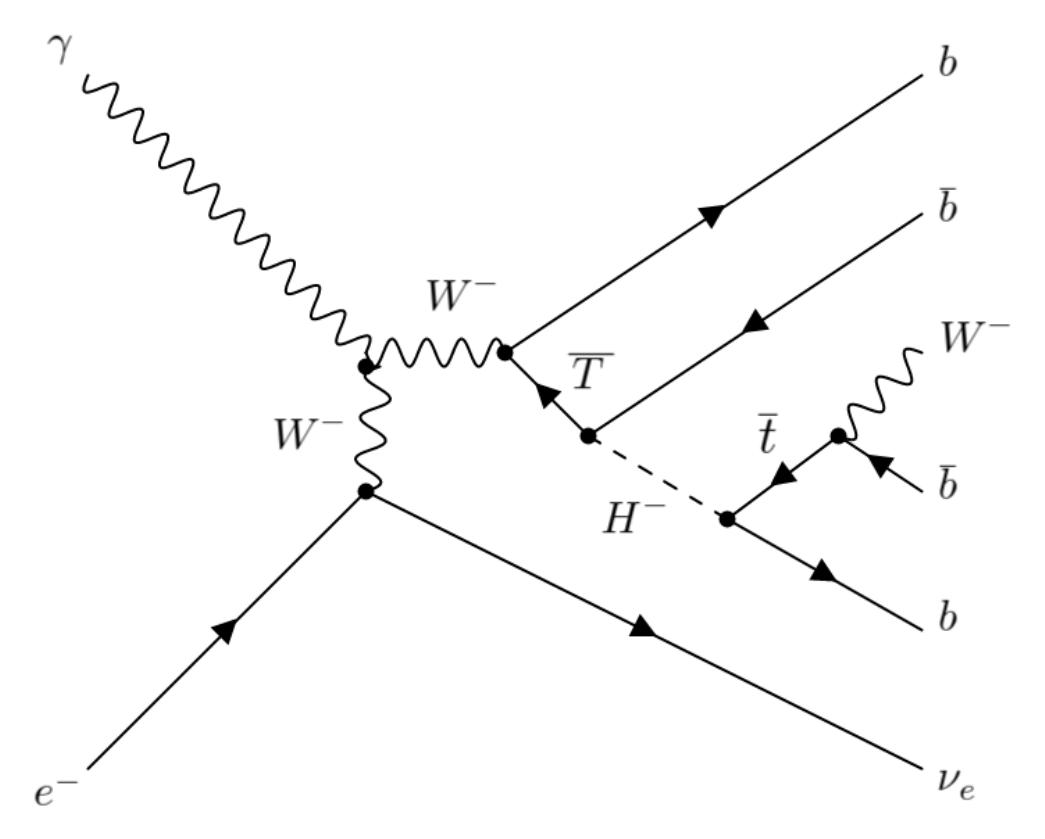}
			\includegraphics[height=3.8cm,width=4cm]{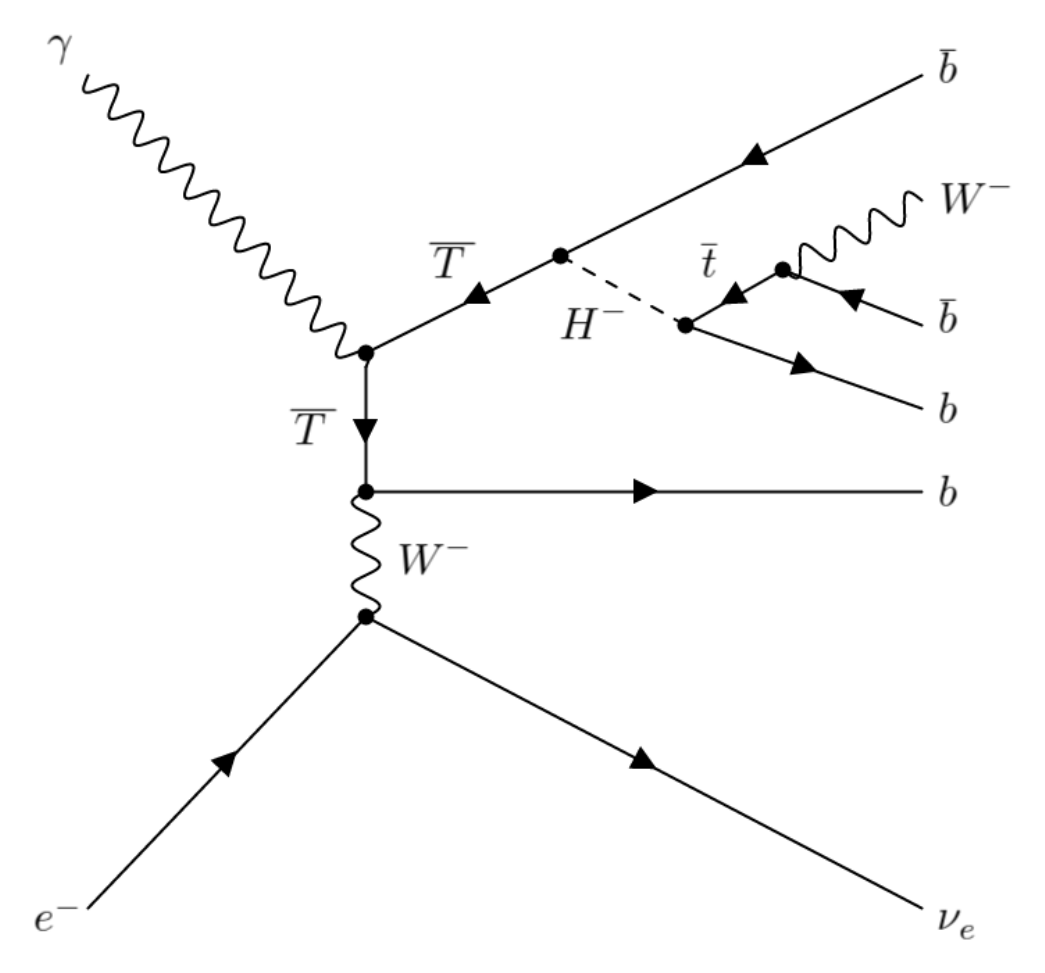}
		\end{minipage}
		\caption{Leading order Feynman diagrams for  $\sigma(e^-\gamma\to \overline{T}b \nu)$ with $T\to H^+b$ and $H^+\to tb$.}
	\end{center}\label{diag}
\end{figure}

The primary SM background processes that could resemble the signal include:

\begin{itemize}
	\item     $\gamma e^- \rightarrow \bar{t}b h\nu _{e},(\bar{t} \rightarrow \bar{b} W^- \rightarrow \bar{b} l^- \nu _l),(h \rightarrow  b \bar{ b})$     
	\item   $\gamma e^- \rightarrow \bar{t}b\nu _{e},(\bar{t} \rightarrow \bar{b} W^- \rightarrow \bar{b} l^- \nu _l)$     
	\item  $\gamma e^- \rightarrow  W^-ZZ\nu _{e},(W^- \rightarrow l^- \nu _l),(Z \rightarrow  b \bar{ b})$     
	\item  $\gamma e^- \rightarrow  W^-Zh\nu _{e},(W^- \rightarrow  l^- \nu _l),(Z,h \rightarrow  b \bar{ b})$   		
	\item  $\gamma e^- \rightarrow  W^-Z\nu _{e},(W^- \rightarrow  l^- \nu _l),(Z \rightarrow  b \bar{ b})$   
	\item $\gamma e^- \rightarrow  W^-h\nu _{e},(W^- \rightarrow  l^- \nu _l),(h \rightarrow  b \bar{ b})$  
	\item  $\gamma e^- \rightarrow \bar{t}bZ\nu _{e},(\bar{t} \rightarrow \bar{b} W^- \rightarrow \bar{b} l^- \nu _l),(Z \rightarrow  b \bar{ b})$    
\end{itemize}

To calculate the cross sections, we first construct the 2HDM-II+$TB$ scenario using the \texttt{FeynRules} package \cite{Degrande:2011ua, Alloul:2013bka} to generate the Universal FeynRules Output (UFO) model files. We then use \texttt{MadGraph5\_aMC\_v2.6.6} \cite{Alwall:2014hca} to compute the leading-order cross sections for the process $e^{-}\gamma \rightarrow \nu_{e} b \bar{T}$. Our analysis focuses on the parameter ranges $m_T \in [500, 2000]$ GeV. The relevant SM parameters are set as follows:
\begin{eqnarray}
&&m_{t}=172.6 ~\mathrm{GeV}, \quad m_{Z}=91.153 ~\mathrm{GeV},\nonumber\\ &&\sin^2(\theta_{W})=0.2226,  \quad\alpha({m_{Z}})= 1/127.934 
\end{eqnarray}

We then assess the impact of beam polarization on the cross sections. Theoretically, the highest cross sections are achieved with full polarization, specifically $P_{e^-} = -1$ and $P_{\gamma} = 1$. However, recognizing the practical polarization limits at facilities like ILC or CLIC where the maximum achievable polarization is approximately $P_{e^-} \times P_{\gamma} = -0.8$ \cite{DeRoeck:2003cjp} we adopt $P_{e^-} = -0.8$ and $P_{\gamma} = 1$ for the polarized case. Fig.~\ref{fig:XS} illustrates the cross sections before the decay of $T$ for both polarized and unpolarized scenarios at center-of-mass energies $\sqrt{s} = 2$ and 3 TeV, with fixed mixing angles $s_R^u=0.05$ and $s_R^d=0.1$. The results clearly demonstrate that polarized beams significantly enhance cross sections, validating their use in this study.

For the simulation, we generate parton-level events for both signal and background processes using \texttt{MadGraph5\_aMC\_v2.6.6}. These events are subsequently processed through \texttt{PYTHIA} \cite{Sjostrand:2014zea} for showering and hadronization. Detector effects are simulated with \texttt{Delphes} \cite{deFavereau:2013fsa} using the appropriate \texttt{ILD} card \cite{ILD1,ILD2}. Jets are clustered using the anti-$k_t$ algorithm \cite{Cacciari:2011ma} with a distance parameter $\Delta R = 0.4$. The final analysis of the signal and background events, post-detector simulation, is performed using \texttt{MadAnalysis} \cite{Conte:2013mea}.

Given the characteristics of our signal, we require at least two b-tagged jets $N(b) \geq 2$ and one lepton $N(l) = 1$. Fig.~\ref{fig2} presents the normalized distributions of key kinematic variables, including the transverse momenta of the leading and subleading b-tagged jets $P_{T}(b_{1})$ and $P_{T}(l)$, the total hadronic transverse energy $H_{T}$, and the separation $\Delta R$ between the lepton and $b_1$. These distributions guide the application of optimized cuts to effectively suppress background contributions.

\begin{figure*}[htpb!]
	\centering
	\includegraphics[height=11.5cm,width=13cm]{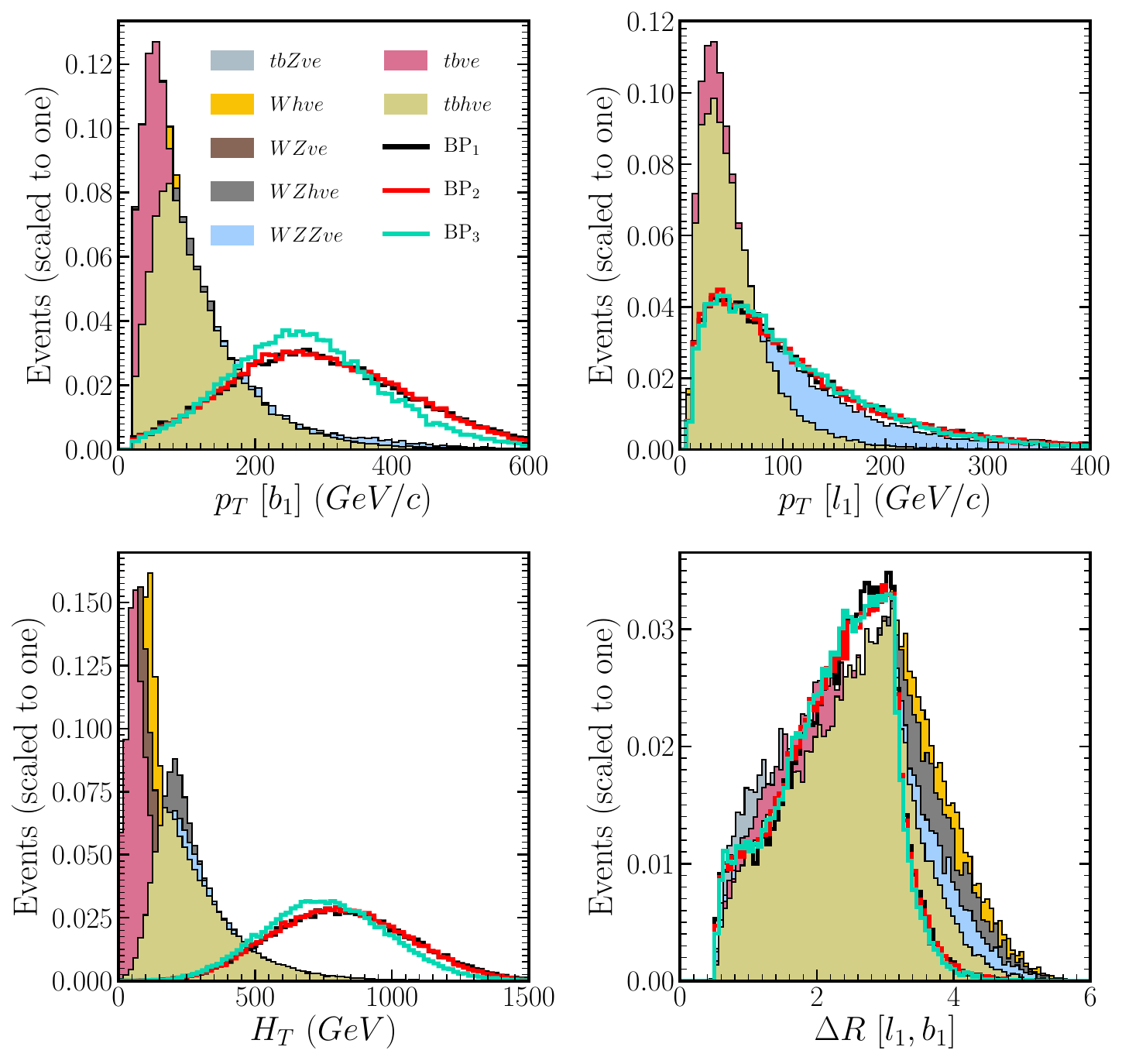}
	\caption{ $p_T(b1)$, $p_T(l)$, $H_T$ and $\Delta R(b_1,~l)$ normalized distributions of the signal BPs (BP$_1$, BP$_2$, BP$_3$) and backgrounds at $\sqrt{s} = 2$ TeV.}
	\label{fig2}
\end{figure*}

To maximize the signal-to-background ratio, we implement a series of optimized cuts informed by the kinematic distributions observed in the simulations. These include $P_{T}(b_{1}) > 160$ GeV, $P_{T}(l) > 30$ GeV, $H_{T} > 650$ GeV, and $\Delta R(l, b_1) < 3.4$. A summary of these cuts is provided in Table \ref{cuts}.

\begin{table}[htp!]
	\centering
	\setlength{\tabcolsep}{15pt}
	\renewcommand{\arraystretch}{1.2}
	\begin{adjustbox}{max width=\textwidth}
		\begin{tabular}{cc} 
			\hline  \hline
			Cut Description & Value \\  
			\hline \hline
			Trigger & $N(l)= 1$, $N(b) \geq 2$ \\  
			\hline
			Cut 1  & $P_{T}^{b_{1}}> 160$ GeV, $P_{T}^{l} > 30$ GeV \\  
			\hline
			Cut 2 & $H_{T}> 650$ GeV \\
			\hline  
			Cut 3 & $\Delta R(b_{1}, l) < 3.4$ \\  
			\hline \hline
		\end{tabular}
	\end{adjustbox}
	\caption{Summary of selection cuts applied to signal and background events.}
	\label{cuts}
\end{table}

Initial state radiation (ISR) can significantly impact cross sections, especially in lepton colliders. Following the methodology outlined in Ref.~\cite{Shang:2019zhh}, we apply a uniform reduction factor of 10\% to account for ISR and beamstrahlung effects on both signal and background cross sections in our significance calculations. Tables \ref{cutflow1} and \ref{cutflow2} provide detailed summaries of the cut-flow for the signal and various backgrounds at $\sqrt{s} = 2$ TeV and $\sqrt{s} = 3$ TeV, respectively, demonstrating the effectiveness of our optimized cuts in distinguishing the signal from the background.

\begin{table*}[htp!]
	\setlength{\tabcolsep}{7.pt}
	\renewcommand{\arraystretch}{1.2}
	\centering %
	\caption{Cut-flow of the cross sections (in $10^{-3}$ fb) for the signals and SM backgrounds at $\sqrt{s}=2$ TeV $e^{-}\gamma$ collider for the three benchmark points. \label{cutflow1}}
	\begin{tabular}{p{1.7cm}<{\centering} p{0.8cm}<{\centering} p{0.8cm}<{\centering} p{0.4cm}<{\centering}p{0.4cm}<{\centering}  p{0.8cm}<{\centering} p{0.8cm}<{\centering} p{1cm}<{\centering} p{1cm}<{\centering} p{0.9cm}<{\centering} p{0.9cm}<{\centering}  p{0.8cm}<{\centering}p{0.8cm}<{\centering}}
		\hline\hline
		\multirow{2}{*}{Cuts}& \multicolumn{3}{c}{Signals }& \multicolumn{7}{c}{ SM~Backgrounds}&  \\ \cline{2-4}  \cline{6-12}
		& BP$_1$ & BP$_2$ & BP$_3 $ & &$tbh\nu$ & $tb\nu$ & $WZZ\nu$&$WZh\nu$&$WZ\nu$&$Whv_{e}$&$tbZ\nu$  \\    \cline{1-8} \hline
		No cuts & 47.7  &  8.18  &  1.62  & & 3.568 & 4372 & 8.17 &5.67& 6108 &6174&5.35 \\
		Trigger& 25.3  &   4.21 & 0.83   && 1.72  & 460& 3.5 &2.57 &953 & 1124&2.32 \\
		Cut 1& 19.6  & 3.26 & 0.64 && 0.33 & 70.4 &0.90 &0.47 & 107 &75.9 & 0.37\\
		Cut 2 & 13.3 & 2.38 &  0.47 && 0.04  & 01.05 &0.071 & 0.027& 0 & 0 & 0.034\\
		Cut 3 & 13.0 & 2.32 &  0.45 && 0.39  & 0.87 &0.069 & 0.027& 0 & 0 & 0.034\\
		
		T.E&27.4$\%$&28.4$\%$&28.2$\%$ &&1.09$\%$&0.02$\%$&0.84$\%$&0.47$\%$&-&-&0.63$\%$\\
		\hline\hline
	\end{tabular}
\end{table*}

\begin{table*}[htp!]
	\setlength{\tabcolsep}{7.pt}
	\renewcommand{\arraystretch}{1.2}
	\centering %
	\caption{The same as in Table~\ref{cutflow1} but for $\sqrt{s}=3$ TeV.\label{cutflow2}}
	\begin{tabular}{p{1.7cm}<{\centering} p{0.8cm}<{\centering} p{0.8cm}<{\centering} p{0.4cm}<{\centering}p{0.4cm}<{\centering}  p{0.8cm}<{\centering} p{0.8cm}<{\centering} p{1cm}<{\centering} p{1cm}<{\centering} p{0.9cm}<{\centering} p{0.9cm}<{\centering}  p{0.8cm}<{\centering}p{0.8cm}<{\centering}}
		\hline\hline
		\multirow{2}{*}{Cuts}& \multicolumn{3}{c}{Signals }& \multicolumn{7}{c}{ SM~Backgrounds}&  \\ \cline{2-4}  \cline{6-12}
		& BP$_1$ & BP$_2$ & BP$_3 $ & &$tbh\nu$ & $tb\nu$ & $WZZ\nu$&$WZh\nu$&$WZ\nu$&$Whv_{e}$&$tbZ\nu$  \\    \cline{1-8} \hline
		No cuts & 415.19  & 131.68   &  40.88  & &8.55  &6592  & 21.02 &15.20 & 10578 &10933 & 11.80 \\
		Trigger& 199   &  61.20  & 18.40   && 3.8  &703 &9.1  & 6.01&1440 &1670 & 4.7\\
		Cut 1& 160  & 50.10 &15.10  && 0.93 &134 & 2.48& 1.42 & 205 & 175 & 1.02 \\
		Cut 2 & 138 & 46.3 & 14.2  && 0.195  & 4.88  & 0.37 &0.14  & 0 & 0 &0.18 \\
		Cut 3 & 135 & 44.90 & 13.70  && 0.17  & 3.56  & 0.32 &0.11  & 0 & 0 &0.16 \\
		
		T.E& 32.6$\%$& 34.1$\%$& 33.5$\%$ &&2.05$\%$&0.05$\%$&1.55$\%$& 0.77$\%$&-&-&1.37$\%$\\
		\hline\hline
	\end{tabular}
\end{table*}

\subsection{Exclusion and Discovery Prospects}
\label{subsec:ExclusionAndDiscovery}

To assess the observability of the signals, we calculate the expected discovery and exclusion significances using the median significance approach as described in Ref.~\cite{Cowan:2010js}. The discovery significance, $\mathcal{Z}_{\mathrm{disc}}$, and exclusion significance, $\mathcal{Z}_{\mathrm{excl}}$, are given by:

\begin{widetext}
	\begin{eqnarray}
	\mathcal{Z}_\mathrm{disc}&=&\sqrt{2\left[\left(s+b\right)\ln\left(\frac{\left(s+b\right)\left(1+\delta^2b\right)}{b+\delta^2b\left(s+b\right)}\right)-\frac{1}{\delta^2}\ln\left(1+\delta^2\frac{s}{1+\delta^2b}\right)\right]}.\\\nonumber
	\mathcal{Z}_\mathrm{excl}&=&\sqrt{2\left[s-b\ln\left(\frac{s+b+x}{2b}\right)-\frac{1}{\delta^2}\ln\left(\frac{b-s+x}{2b}\right)\right]-\left(b+s-x\right)\left(1+\frac{1}{\delta^2b}\right)}.
	\end{eqnarray}
\end{widetext}
where
\begin{equation*}
x=\sqrt{\left(s+b\right)^2-4\delta^2sb^2/\left(1+\delta^2b\right)}.
\end{equation*}

In the limit of $\delta\to0$, these expressions simplify to:
\begin{eqnarray}
\mathcal{Z}_\mathrm{disc}&=&\sqrt{2\left[\left(s+b\right)\ln\left(1+s/b\right)-s\right]},\\\nonumber
\mathcal{Z}_\mathrm{excl}&=&\sqrt{2\left[s-b\ln\left(1+s/b\right)\right]}.
\end{eqnarray}
Here, $s$ and $b$ represent the expected number of signal and total SM background events after all selection cuts, while $\delta$ denotes the systematic uncertainty. Although the primary analysis considers a systematic uncertainty of $\delta = 10\%$, Tables \ref{Tab_sig2TeV} and \ref{Tab_sig3TeV} also present discovery significances for various VLT masses $m_T$, integrated luminosities (1500 fb$^{-1}$ and 3000 fb$^{-1}$), and systematic uncertainties ($\delta = 0\%$, 10\%, 30\%, 50\%) at center-of-mass energies of $\sqrt{s} = 2$ TeV and $\sqrt{s} = 3$ TeV.

The results reveal that at $\sqrt{s} = 2$ TeV, the expected significance decreases as the VLT mass increases, especially under higher systematic uncertainties. For an integrated luminosity of 1500 fb$^{-1}$, the significance reaches 3.67 for a 1000 GeV VLT mass with $\delta = 0\%$, indicating evidence for the signal. However, as the mass increases to 1600 GeV, the significance diminishes to 0.20, falling well below the threshold for evidence or discovery. Even with an increased luminosity of 3000 fb$^{-1}$, the significance improves reaching a maximum of 5.19 for 1000 GeV but continues to reduce with increasing mass, and higher systematic uncertainties further impact the discovery potential.

In contrast, at $\sqrt{s} = 3$ TeV, the discovery potential is significantly enhanced. For a 1000 GeV VLT mass, the significance at 1500 fb$^{-1}$ reaches 15.47 with $\delta = 0\%$, far exceeding the discovery threshold. Even with systematic uncertainties as high as $\delta = 50\%$, the significance remains robust at 7.05. As the VLT mass increases to 1600 GeV, the significance decreases but remains substantial, particularly with 3000 fb$^{-1}$, where it reaches 5.53 for $\delta = 0\%$, indicating a potential discovery. These results clearly demonstrate that higher center-of-mass energy and increased luminosity greatly enhance the sensitivity to VLT searches, particularly for heavier masses, even in the presence of significant systematic uncertainties.
\begin{table*}[ht!]
	\centering
	\includegraphics[height=4.8cm,width=14cm]{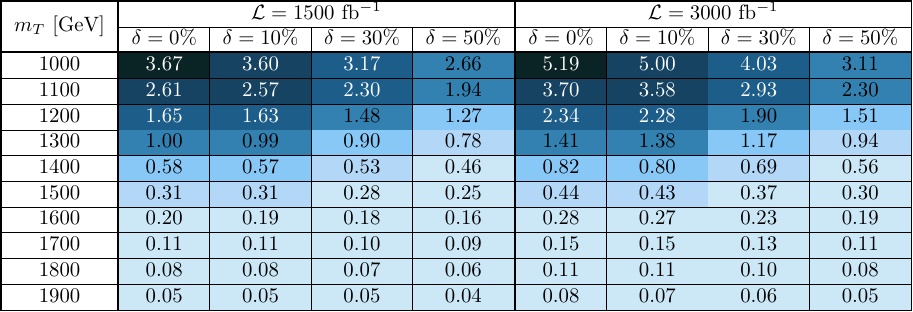}
	\caption{Discovery significance, $\mathcal{Z}\mathrm{disc}$, for $\sqrt{s} = 2$ TeV, presented at varying systematic uncertainties ($\delta$) and integrated luminosities ($\mathcal{L} = 1500$ fb$^{-1}$ and $\mathcal{L} = 3000$ fb$^{-1}$). The parameters are fixed as follows: $m_H = 600.26$ GeV, $m_A = 595.24$ GeV, $m_{H^\pm} = 658.07$ GeV, $\tan\beta = 6$, $s^u_R = 0.05$, and $s^d_R = 0.11$. All points depicted are consistent with the discussed constraints.}
	\label{Tab_sig2TeV}
\end{table*}
\begin{table*}[ht!]
	\centering
	\includegraphics[height=5cm,width=14cm]{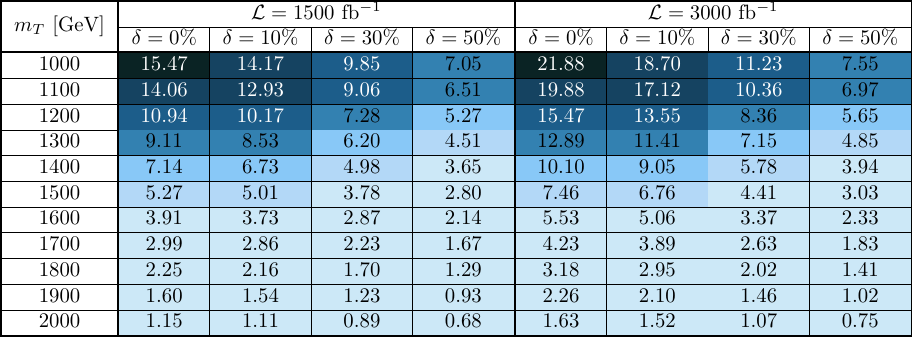}
	\caption{The same as in Table~\ref{Tab_sig2TeV} but for $\sqrt{s}=3$ TeV.}
	\label{Tab_sig3TeV}
\end{table*}

\begin{figure*}[htp!]
	\centering
	\includegraphics[height=7.cm,width=7.5cm]{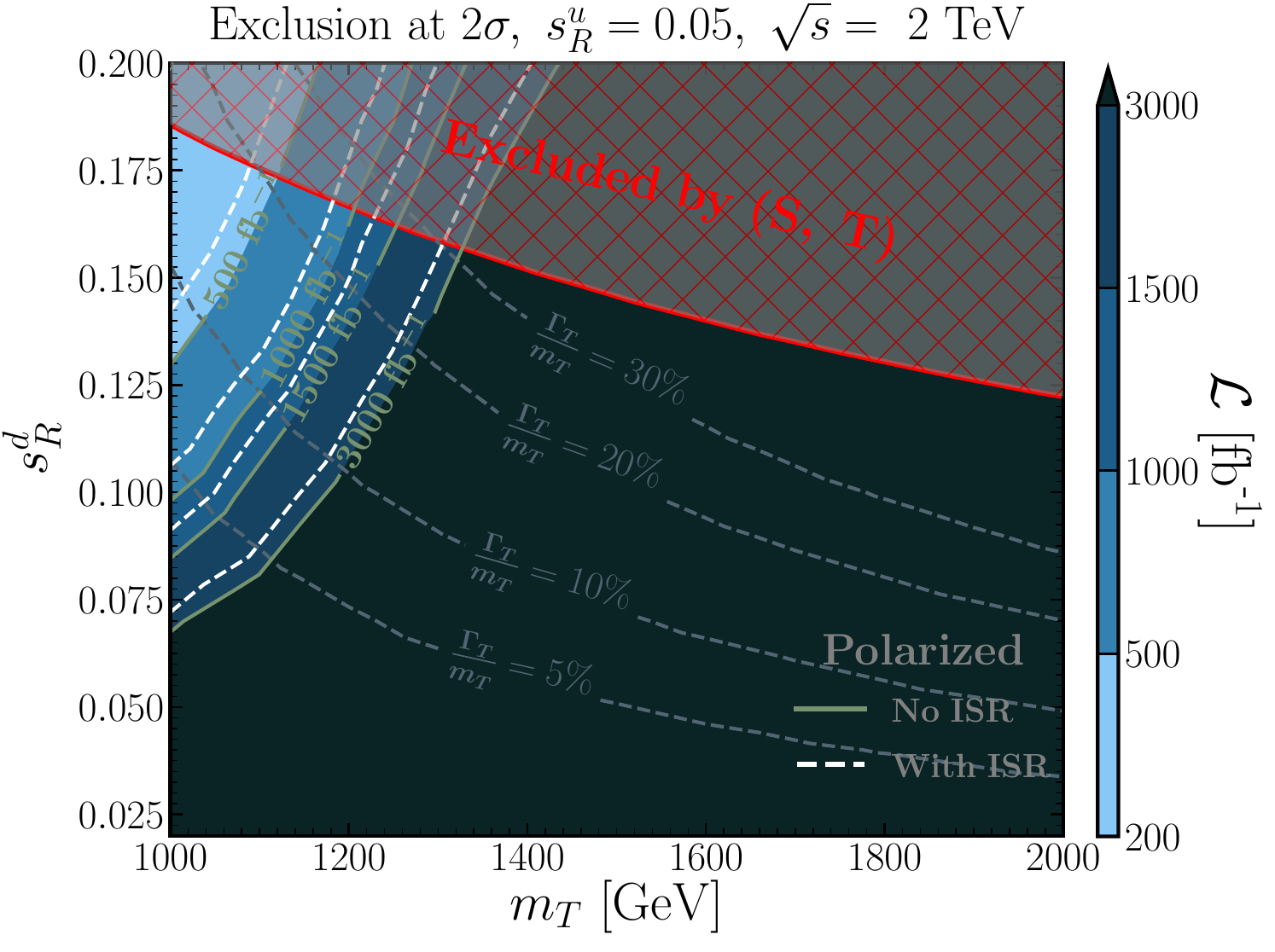}
	\includegraphics[height=7.cm,width=7.5cm]{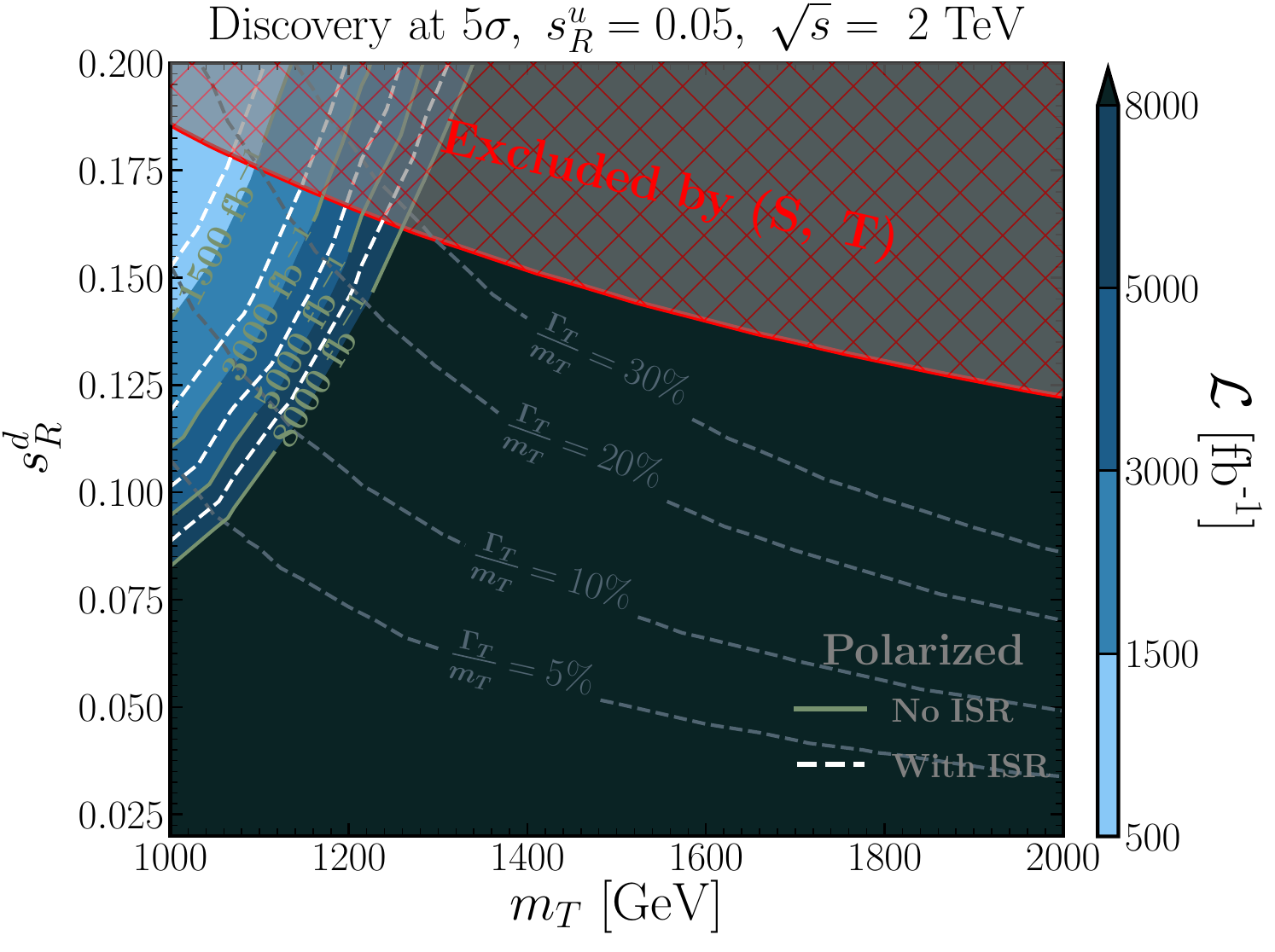}
	\caption{The discovery prospects (5$\sigma$) and exclusion limits (2$\sigma$) for the signal in $m_T-s^d_R$ planes at $\sqrt{s}=2$ TeV ILC for different integrated luminosities. The results are shown both with and without ISR and beamstrahlung effects, with fixed parameters: $m_H=600.26$ GeV, $m_A=	595.24$ GeV, $m_{H^\pm}= 658.07$ GeV, $\tan\beta=6$ and $s^u_R=0.05$.}
	\label{fig5}
\end{figure*}

\begin{figure*}[htp!]
	\centering
	\includegraphics[height=7.cm,width=7.5cm]{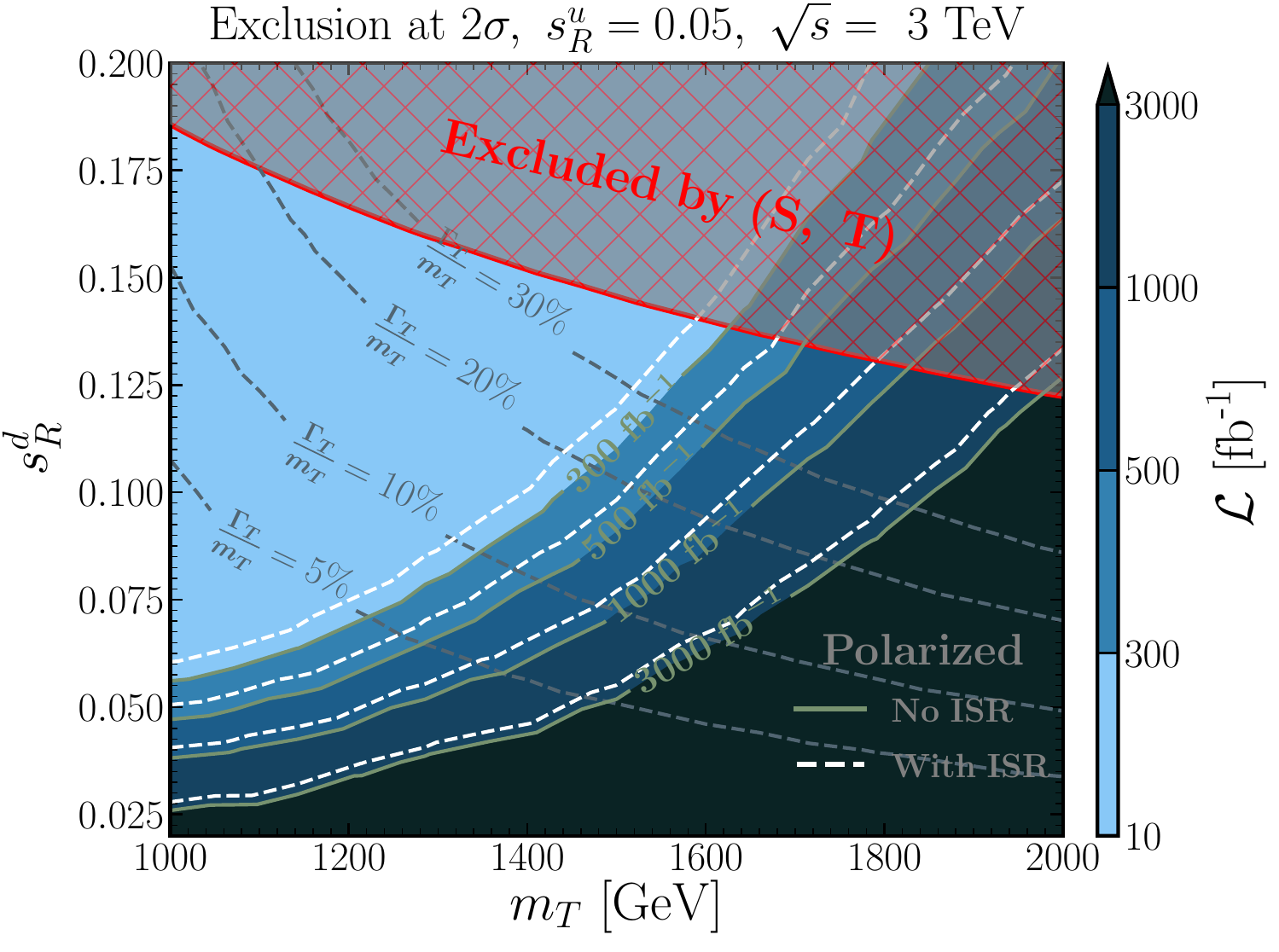}
	\includegraphics[height=7.cm,width=7.5cm]{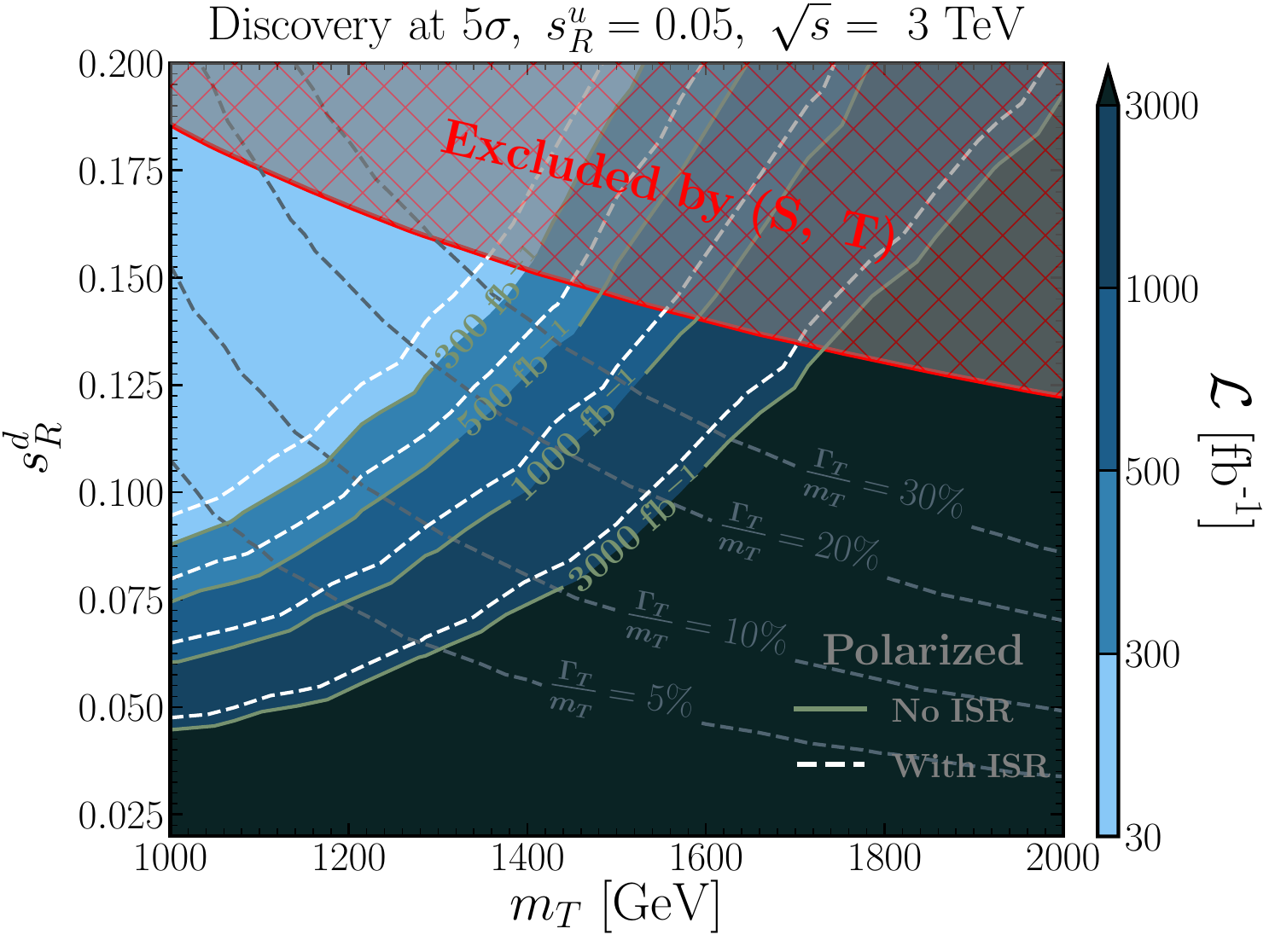}
	\caption{The same as in Fig.~\ref{fig5} but for $\sqrt{s}=3$ TeV.}
	\label{fig6}
\end{figure*}

Before concluding, we present the 2$\sigma$ exclusion limits (left panels) and 5$\sigma$ discovery potentials (right panels) in the $(m_T, s^d_R)$ plane as shown in Figs.~\ref{fig5} and \ref{fig6}, for center-of-mass energies $\sqrt{s}=2$ and 3 TeV, respectively, under polarized beam conditions with a background uncertainty of $\delta=10\%$. The red hatched regions indicate areas excluded by EWPOs ($S$, $T$). Solid lines correspond to different integrated luminosities, while dashed white lines illustrate the impact of ISR effects. Grey dashed lines represent the width-to-mass ratios $\Gamma_T/m_T$ up to 30\%.

The results demonstrate that EWPOs impose stringent constraints on large mixing angles, thereby reducing the parameter space for exclusion by current LHC limits from searches for singly produced VLTs.
Focusing on the regions allowed by EWPOs, at $\sqrt{s}=2$ TeV with an integrated luminosity of 1500 fb$^{-1}$, the excluded region is $s_R^d \in [0.085, 0.16]$ for $m_T \in [1000, 1260]$ GeV, while the discovery region is $s_R^d \in [0.14, 0.17]$ for $m_T \in [1000, 1100]$ GeV. Increasing the luminosity to 3000 fb$^{-1}$ shifts the exclusion region to $s_R^d \in [0.05, 0.15]$ for $m_T \in [1000, 1340]$ GeV and the discovery region to $s_R^d \in [0.11, 0.17]$ for $m_T \in [1000, 1160]$ GeV. The inclusion of ISR effects further reduces both exclusion and discovery capabilities.
At $\sqrt{s}=3$ TeV, with an integrated luminosity of 300 fb$^{-1}$, the excluded region is $s_R^d \in [0.055, 0.135]$ for $m_T \in [1000, 1640]$ GeV, and the discovery region is $s_R^d \in [0.09, 0.15]$ for $m_T \in [1000, 1400]$ GeV. With a luminosity of 1000 fb$^{-1}$, the exclusion region becomes $s_R^d \in [0.045, 0.125]$ for $m_T \in [1000, 1820]$ GeV, and the discovery region is $s_R^d \in [0.058, 0.14]$ for $m_T \in [1000, 1600]$ GeV. At 3000 fb$^{-1}$, the exclusion range narrows to $s_R^d \in [0.028, 0.12]$ for $m_T \in [1000, 1970]$ GeV, while the discovery region expands to $s_R^d \in [0.04, 0.122]$ for $m_T \in [1000, 1760]$ GeV. ISR effects consistently diminish both exclusion and discovery potential. Overall, the discovery potential is significantly enhanced at $\sqrt{s}=3$ TeV compared to $\sqrt{s}=2$ TeV.

\section{SUMMARY}
\label{sec:conclusion}

In this study, we investigated the production and decay of the vector-like top quark (VLT) within the 2HDM Type II framework, extended by a $TB$ doublet, with a focus on $e^{-}\gamma$ collisions at future linear colliders. The decay channel $\bar{T} \rightarrow H^{-} \bar{b}$ was identified as the dominant mode for $m_T$ values above 800 GeV, forming a robust basis for further exploration. Through an in-depth analysis of the parameter space, three benchmark points were selected, satisfying all relevant theoretical and experimental constraints.

\noindent We computed leading-order cross sections for the process $e^{-}\gamma \rightarrow \nu_e b \bar{T}$ under both polarized and unpolarized beam conditions, showing that polarized beams significantly enhance detection prospects. Detector effects were modeled, and optimized selection criteria were applied to maximize signal significance while minimizing background interference. The analysis included the impact of initial state radiation (ISR) and beamstrahlung effects, ensuring a comprehensive evaluation of signal viability.

\noindent Our study assessed the exclusion and discovery prospects for the VLT and the charged Higgs boson $H^{\pm}$ at center-of-mass energies of 2 and 3 TeV across varying integrated luminosities. The results indicate that future linear colliders, such as the ILC, could effectively probe the extended scalar sector and vector-like quark signatures within the 2HDM + VLQ framework, even accounting for ISR and beamstrahlung effects.

\bibliography{main}
\bibliographystyle{apsrev4-2}
\end{document}